\documentclass[12pt]{article}
\usepackage{graphicx}
\usepackage{amsmath,amssymb}
\usepackage{hyperref}

\newcommand\pubnumber{CIPANP2018-Watanabe}
\newcommand\pubdate{\today}

\def\napoli{Physique des Particules, Universit\'e de Montr\'eal, \\ C.P. 6128, succ.\ centre-ville, Montr\'eal, QC, Canada H3C 3J7}

\def\Title#1{\begin{center} {\Large #1 } \end{center}}
\def\Author#1{\begin{center}{ \sc #1} \end{center}}
\def\Address#1{\begin{center}{ \it #1} \end{center}}

\newcommand\pubblock{\rightline{\begin{tabular}{l} \pubnumber\\
         \pubdate  \end{tabular}}}
\newenvironment{Abstract}{\begin{quotation}  }{\end{quotation}}
\newenvironment{Presented}{\begin{quotation} \begin{center} 
             PRESENTED AT\end{center}\bigskip 
      \begin{center}\begin{large}}{\end{large}\end{center} \end{quotation}}
\def\Acknowledgements{\bigskip  \bigskip \begin{center} \begin{large}
             \bf ACKNOWLEDGEMENTS \end{large}\end{center}}

\def\beq{\begin{equation}}
\def\eeq#1{\label{#1}\end{equation}}
\def\eeqn{\end{equation}}

\def\beqa{\begin{eqnarray}}
\def\eeqa#1{\label{#1}\end{eqnarray}}
\def\eeqan{\end{eqnarray}}

\let\bar=\overbar

\def\Dslash{\not{\hbox{\kern-4pt $D$}}}
\def\dslash{\not{\hbox{\kern-2pt $\del$}}}

\def\msb{{\bar{\ssstyle M \kern -1pt S}}}

\begin{document}
\begin{titlepage}
\pubblock

\vfill
\Title{$R_D$ and $R_{D^*}$: Theoretical Development}
\vfill
\Author{ Ryoutaro Watanabe}
\Address{\napoli}
\vfill
\begin{Abstract}
Recent theoretical developments on $R_D$ and $R_{D^*}$ -- discrepancies between experimental data and the Standard Model predictions have been reported (B anomaly) -- are reviewed. 
New Physics explanations for the B anomaly and other relevant observables to obtain additional bounds on New Physics are also summarized. 
Note: this is the proceedings for the talk at CIPANP2018 which has been held on May 29 2018, and thus quite recent works are not mentioned. 
\end{Abstract}
\vfill
\begin{Presented}
CIPANP2018 \\[0.5em]
Palm Springs, California, May 29 -- June 3, 2018
\end{Presented}
\vfill
\end{titlepage}
\def\thefootnote{\fnsymbol{footnote}}
\setcounter{footnote}{0}

\section{Introduction}

\begin{figure}[h!]
\centering
\includegraphics[height=2.5in]{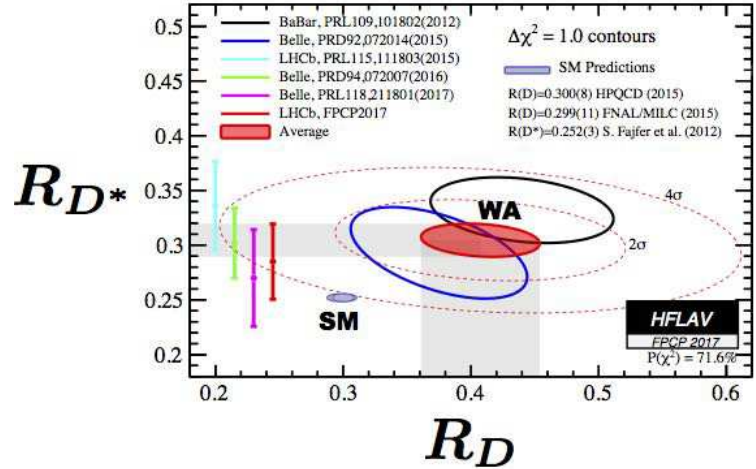}
\caption{Present status of the anomalies in $R_{D^{(*)}}$.
The world average of experimental data has been evaluated in Ref.~\cite{HFLAV}. }
\label{fig:HFLAV}
\end{figure}

The observed excesses in $\bar B \to D \tau\bar\nu$ and $\bar B \to D^* \tau\bar\nu$ have been one of the major anomalies in particle physics. 
These exclusive modes are predicted at the tree level in the standard model (SM), and thus large discrepancies with experimental data immediately implies existence of new physics (NP) in case that it is statistically confirmed. 
For now, the anomalies have been observed in the ratios to light leptonic modes ($\ell = e,\mu$), defined as 
\begin{align}
 R_D = \frac{\Gamma (\bar B \to D \tau\bar\nu)}{\Gamma (\bar B \to D \ell\bar\nu)} \,,
 ~~~
 R_{D^*} = \frac{\Gamma (\bar B \to D^* \tau\bar\nu)}{\Gamma (\bar B \to D^* \ell\bar\nu)} \,. 
\end{align}
Following the report by HFLAV~\cite{HFLAV}, the current situation for both the SM predictions and the experimental data is summarized as in Fig.~\ref{fig:HFLAV}. 
The discrepancy reaches $4\sigma$ level at present.

In this paper, we briefly review recent works trying to evaluate precise values for the SM predictions on $R_D$ and $R_{D^*}$ and also summarize possibilities of NP explanations to the present anomalies. 
Finally we will mention other relevant observables which can probe and/or distinguish the NP effects.

\section{SM predictions}

The SM predictions on the branching ratios of $\bar B \to D^{(*)} \tau\bar\nu$ have been done by several works. 
Usually, heavy quark effective theory (HQET) has been applied to parametrize form factors of the $\bar B \to D^{(*)}$ transitions with the expansions of $\Lambda_\text{QCD}/M_Q$ and $\alpha_s$, 
based on the Caprini-Lellouch-Neubert (CLN) parameterization~\cite{Caprini:1997mu}. 
The form factors included in the light leptonic modes $\bar B \to D^{(*)} \ell\bar\nu$ are then determined by fits to the experimental data 
while those that only appear in the tauonic modes have to be evaluated with lattice QCD study, ({\it e.g.}, see Ref.~\cite{Tanaka:2012nw}.) 
Recently there are two developments on the evaluations of the form factors.

In Refs.~\cite{Bigi:2016mdz,Bigi:2017jbd}, the authors pointed out that the Boyd-Grinstein-Lebed parameterization~\cite{Boyd:1997kz} for the form factors in $\bar B \to D \ell\bar\nu$ could provide a more precise fit to data 
in case that a large amount of signal events is available, such as the Belle~II experiment.
In Ref.~\cite{Bernlochner:2017jka}, the authors have included $\mathcal O(\Lambda_\text{QCD}/M_Q)$ and $\mathcal O(\alpha_s)$ contributions to the decay rates, 
which were previously taken as parts of the theoretical uncertainties in the CLN parameterization. 
With the modified parametrizations taking these contributions into account, combined fits to the experimental data of $\bar B \to D^{(*)} \ell\bar\nu$ have been performed. 
Then these works have obtained the SM predictions as 
\begin{align}
 R_D^\text{SM} = 0.299 \pm 0.003 \,,
 ~~~
 R_{D^*}^\text{SM} = 0.257 \pm 0.003  \,. 
\end{align}
Therefore, the uncertainty is at 1\% level for now.

\begin{figure}[h!]
\centering
\includegraphics[height=2.5in]{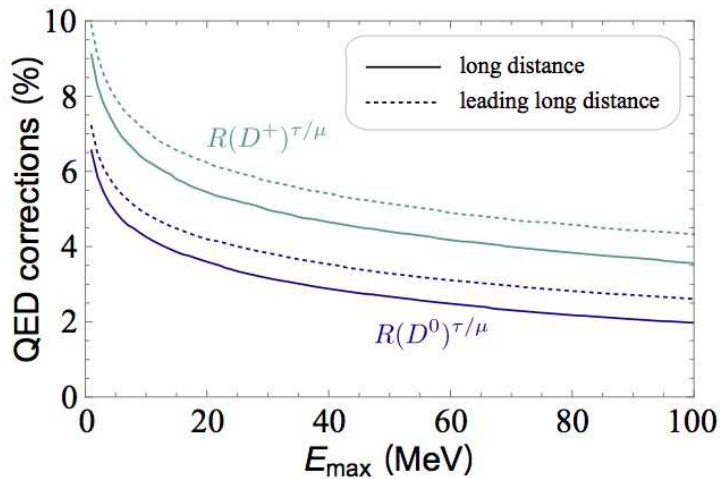}
\caption{
The soft-photon corrections to $R_{D^+}$ and $R_{D^0}$ as a function of the photon energy cut. 
For the detail, see Ref.~\cite{deBoer:2018ipi}.  
}
\label{fig:QED}
\end{figure}

Another development was given in the work of Ref.~\cite{deBoer:2018ipi} that evaluates soft-photon effects on $R_D$. 
Significant points are summarized as follows. 
(1) The effect of soft-photon emission depends on the lepton mass and thus difference between $\bar B \to D \tau\bar\nu$ and $\bar B \to D \ell\bar\nu$ leads to a correction on $R_D$. 
(2) Photons are emitted from charged particles, which implies that $\bar{B}^0 \to D^+$ and $B^- \to D^0$ transitions have different effects and also depend on photon energy cut. 
The net effect gives non-negligible constructive contribution to $R_D$ at $\sim 5\%$ level, as shown in Fig.~\ref{fig:QED}.

\section{NP explanations}

Here we employ the effective Lagrangian for $b \to c \tau\bar\nu_\tau$ to investigate NP explanations, defined as 
\begin{align}
 -\mathcal L = 2\sqrt 2 G_F V_{cb} C_\text{NP} \mathcal O_\text{NP} \,, 
\end{align}
where $C_\text{NP}$ is a Wilson coefficient for a corresponding NP operator $\mathcal O_\text{NP}$. 
For example, when we introduce $\mathcal O_{V_1} \equiv (\bar c \gamma^\mu P_L b) (\bar\tau \gamma_\mu P_L\nu_\tau)$, we can obtain a fit for $C_{V_1}$ to the present experimental data.

In Table~\ref{tab:NPsolution}, we summarize NP solutions to the $R_{D^{(*)}}$ anomalies. 
For the $V_1$ operator, a 17\% NP contribution of the SM value ($C_{V_1}^\text{SM}=1$) is necessary to accommodate the central values of the present experimental data. 
Assuming a tree level NP interaction with couplings $=1$, this result implies existence of $\sim 2\,\text{TeV}$ scale NP. 
For the $V_2$ operator, a complex number of the Wilson coefficient is required as the best fit solution, $C_{V_2} \sim 0.01 + 0.6 i$. 
As for the scalar operators, the $S_1$ scenario -- corresponding to the main contribution in Two Higgs Doublet Model (THDM) of type~II -- is disfavored (no solution to the best fit values).
On the other hand, the $S_2$ scenario has a solution, but needs the large distractive contribution, $C_{S_2} \sim -1.5$. 
Finally, the tensor type operator can explain the data with $C_{T} \sim 0.3$.

\begin{table}[h!]
\begin{center}
\begin{tabular}{cc|c} 
\hline\hline
Operator &  Best fit solution & NP example \\
\hline
$\mathcal O_{V_1} \equiv (\bar c \gamma^\mu P_L b) (\bar\tau \gamma_\mu P_L\nu_\tau)$  &  $C_{V_1} \sim 0.17$  &  $W'$, Vector leptoquark \\
\hline
$\mathcal O_{V_2} \equiv (\bar c \gamma^\mu P_R b) (\bar\tau \gamma_\mu P_L\nu_\tau)$  &  $C_{V_2} \sim 0.01 + 0.6 i$  &  $W'$ \\
\hline
$\mathcal O_{S_1} \equiv (\bar c P_R b) (\bar\tau P_L\nu_\tau)$  &  no solution & Charged Higgs, Scalar leptoquark \\
\hline
$\mathcal O_{S_2} \equiv (\bar c P_L b) (\bar\tau P_L\nu_\tau)$  &  $C_{S_2} \sim -1.5$ & Charged Higgs \\
\hline
$\mathcal O_{T} \equiv (\bar c \sigma^{\mu\nu} P_L b) (\bar\tau \sigma_{\mu\nu} P_L\nu_\tau)$  &  $C_{T} \sim 0.3$ & Doublet vector/scalar leptoquark \\
\hline\hline
\end{tabular}
\caption{NP solutions to the central experimental values of the present data of $R_D$ and $R_{D^*}$ in terms of the Wilson coefficient. 
Example of NP particles is also shown. }
\label{tab:NPsolution}
\end{center}
\end{table}

\section{Relevant observables}

The integrated decay rates of $\bar B \to D^{(*)} \tau\bar\nu$, equivalently to $R_{D^{(*)}}$, have been well analyzed with the data collected at BaBar, Belle, and LHCb. 
We expect that a large amount of signal events enables us to investigate distribution(s) of the processes at the future Belle~II experiment. 
In light of this, the work in Ref.~\cite{Sakaki:2014sea} has estimated statistics obtained at Belle~II for the $q^2 = (p_B-p_{D^{(*)}})^2$ distribution 
and then it turns out that we can distinguish the above NP scenarios of Table~\ref{tab:NPsolution} if we use the $q^2$ distributions of $5\,\text{ab}^{-1}$ data at Belle~II. 
For details in future projections with the use of distributions, see Ref.~\cite{Kou:2018nap}.

\begin{figure}[h!]
\centering
\includegraphics[height=2.0in]{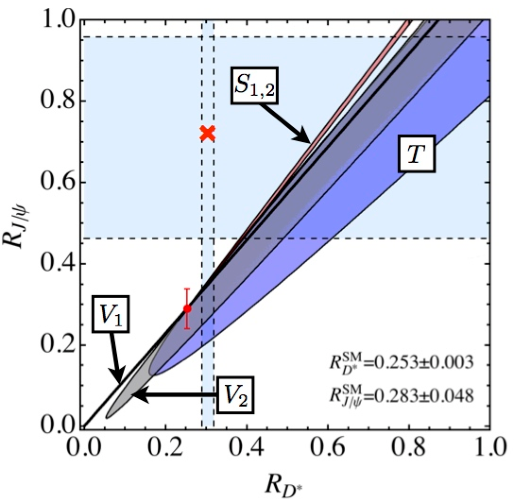}~~~
\includegraphics[height=2.0in]{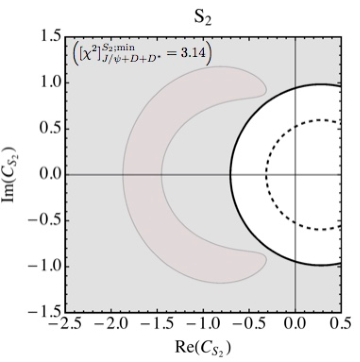}
\caption{
Left: correlation between $R_{D^*}$ and $R_{J/\psi}$ in the presence of one NP operator, $V_{1,2}$, $S_{1,2}$, or $T$. 
Right: allowed region of $C_{S_2}$ from the $R_{D^{(*)}}$ anomalies [red] and the indirect $B_c$ bound [gray].  }
\label{fig:Jpsi}
\end{figure}

Observables other than $\bar B \to D^{(*)} \tau\bar\nu$ have turned out to be significant as several developments have been reported. 
In Ref.~\cite{Watanabe:2017mip}, $B_c \to J\!/\!\psi\, \tau\nu$ has been considered to test NP along with the $R_{D^{(*)}}$ anomalies, 
as this process has been observed at LHCb~\cite{Aaij:2017tyk} in the ratio of 
\begin{align}
 R_{J/\psi} = \frac{\Gamma (B_c \to J\!/\!\psi\, \tau\bar\nu)}{\Gamma (B_c \to J\!/\!\psi\, \mu\bar\nu)} \,. 
\end{align}
The SM prediction and data are then $R_{J/\psi}^\text{SM} = 0.283 \pm 0.048$ and $R_{J/\psi}^\text{LHCb} = 0.71 \pm 0.17 \pm 0.18$, respectively. 
Thus, one finds that there is a $1.7\sigma$ deviation although the experimental data still includes a large error. 
In the left panel of Fig.~\ref{fig:Jpsi}, correlation between $R_{D^*}$ and $R_{J/\psi}$ in the presence of one NP operator is shown. 
Thus we can see that the NP solution to the $R_{D^{(*)}}$ anomalies, as given in Table~\ref{tab:NPsolution}, is not compatible with the present $R_{J/\psi}$ observation.

In Refs.~\cite{Alonso:2016oyd,Akeroyd:2017mhr}, $B_c \to \tau\nu$ has been used to indirectly obtain NP bound in $b \to c \tau\bar\nu$. 
The indirect constraints come from the $B_c$ lifetime~\cite{Alonso:2016oyd} and LEP data extracted from $Z$ boson peak~\cite{Akeroyd:2017mhr}, 
which leads to the limit as $\mathcal B (B_c \to \tau\nu) < 10 \,\text{--}\, 30 \%$. 
This indirect bound is superimposed on the allowed region for the $S_2$ scenario in the right panel of Fig.~\ref{fig:Jpsi}. 
Therefore, we can conclude that the $S_2$ solution to the $R_{D^{(*)}}$ anomalies is excluded.

\section{Summary}

We have reviewed recent theoretical developments on $R_D$ and $R_{D^*}$, and on relevant observables 
such as the $q^2$ distributions of $\bar B \to D^{(*)} \tau\bar\nu$, the ratio of $B_c \to J\!/\!\psi\, \tau\nu$ / $B_c \to J\!/\!\psi\, \mu\nu$, and the indirect bound from the $B_c$ decay.

The SM predictions have been revisited with the dedicated studies on the form factors of the $\bar B \to D^{(*)}$ transitions for the purpose of precise evaluation, 
which results in 1\% level uncertainties in $R_D^\text{SM}$ and $R_{D^*}^\text{SM}$. 
It is another development that the soft-photon effects on $R_D$ -- long distance correction from QED -- have been newly evaluated. 
Then it was found that the corrections to the ratios are not negligible but at 5\% level constructive to the tree level SM values.

NP explanations to the central values of the present experimental data are summarized in terms of the Wilson coefficient for the possible NP operators. 
It turned out that there exist solutions to the $R_{D^{(*)}}$ anomalies in the vector, scalar, and tensor types of NP interactions.

Relevant observables other than $R_{D^{(*)}}$ have been pointed out to further constrain NP interactions in $b \to c \tau\bar\nu$. 
The $q^2$ distributions of $\bar B \to D^{(*)} \tau\bar\nu$ have been investigated to estimate its potential reachable at the Belle~II experiment. 
Then it was found that $5\,\text{ab}^{-1}$ of accumulated data at Belle~II is sufficient to probe and/or distinguish the NP solutions to the present data of $R_{D^{(*)}}$. 
The other observations for $B_c$ have been used to provide additional bounds on NP. 
The ratio $R_{J/\psi}$ ($B_c \to J\!/\!\psi\, \tau\nu$ / $B_c \to J\!/\!\psi\, \mu\nu$) was observed at LHCb whose result is not consistent with the SM prediction at $1.7\sigma$. 
Its central value cannot be accommodated with the NP solutions to the $R_{D^{(*)}}$ anomalies although the data still includes a large error. 
The $B_c$ lifetime gives the indirect limit on NP interactions.
In particular, the scalar type NP interaction contributes to $B_c \to \tau\nu$ significantly and indeed it excludes the NP solution with the scalar type interaction.

These theoretical developments will be more significant when the Belle~II experiment starts to accumulate signal events for $\bar B \to D^{(*)} \tau\bar\nu$. 
Also, larger amounts of accumulated data will enable us to utilize multiple distributions, from which tau polarization and some angular asymmetries can be extracted, for further details of NP investigation
({\it e.g.}, see Refs.~\cite{Sakaki:2013bfa,Alonso:2017ktd}.) 
A summary on the Belle~II physics from both theoretical and experimental sides is now available in Ref.~\cite{Kou:2018nap}.

\Acknowledgements
I am grad to be one of the conveners in The Belle II Theory Interface Platform working groups (B2TiP) that has published The Belle II Physics Book~\cite{Kou:2018nap}. 
I am grateful to Mugi Ienaga for her healing hospitality while I was having misfortune.


\begin{thebibliography}{99}


\bibitem{HFLAV}
  \href{https://hflav-eos.web.cern.ch/hflav-eos/semi/summer18/RDRDs.html}{{\ttfamily https://hflav-eos.web.cern.ch/hflav-eos/semi/summer18/RDRDs.html}}

\bibitem{Caprini:1997mu} 
  I.~Caprini, L.~Lellouch and M.~Neubert,
  Nucl.\ Phys.\ B {\bf 530}, 153 (1998) 
  [hep-ph/9712417].

\bibitem{Tanaka:2012nw} 
  M.~Tanaka and R.~Watanabe,
  Phys.\ Rev.\ D {\bf 87}, no. 3, 034028 (2013) 
  [arXiv:1212.1878 [hep-ph]].

\bibitem{Bigi:2016mdz} 
  D.~Bigi and P.~Gambino,
  Phys.\ Rev.\ D {\bf 94}, no. 9, 094008 (2016)
  doi:10.1103/PhysRevD.94.094008
  [arXiv:1606.08030 [hep-ph]].
  
\bibitem{Bigi:2017jbd} 
  D.~Bigi, P.~Gambino and S.~Schacht,
  JHEP {\bf 1711}, 061 (2017)
  doi:10.1007/JHEP11(2017)061
  [arXiv:1707.09509 [hep-ph]].

\bibitem{Boyd:1997kz} 
  C.~G.~Boyd, B.~Grinstein and R.~F.~Lebed,
  Phys.\ Rev.\ D {\bf 56}, 6895 (1997)
  [hep-ph/9705252].
  
\bibitem{Bernlochner:2017jka} 
  F.~U.~Bernlochner, Z.~Ligeti, M.~Papucci and D.~J.~Robinson,
  Phys.\ Rev.\ D {\bf 95}, no. 11, 115008 (2017)
  Erratum: [Phys.\ Rev.\ D {\bf 97}, no. 5, 059902 (2018)]~
  [arXiv:1703.05330 [hep-ph]].

\bibitem{deBoer:2018ipi} 
  S.~de Boer, T.~Kitahara and I.~Nisandzic,
  Phys.\ Rev.\ Lett.\  {\bf 120}, no. 26, 261804 (2018)
  [arXiv:1803.05881 [hep-ph]].
  
\bibitem{Sakaki:2014sea} 
  Y.~Sakaki, M.~Tanaka, A.~Tayduganov and R.~Watanabe,
  Phys.\ Rev.\ D {\bf 91}, no. 11, 114028 (2015)
  [arXiv:1412.3761 [hep-ph]].  

\bibitem{Kou:2018nap} 
  E.~Kou {\it et al.} [Belle II Collaboration],
  arXiv:1808.10567 [hep-ex].

\bibitem{Watanabe:2017mip} 
  R.~Watanabe,
  Phys.\ Lett.\ B {\bf 776}, 5 (2018)
  [arXiv:1709.08644 [hep-ph]].

\bibitem{Aaij:2017tyk} 
  R.~Aaij {\it et al.} [LHCb Collaboration],
  Phys.\ Rev.\ Lett.\  {\bf 120}, no. 12, 121801 (2018)
  [arXiv:1711.05623 [hep-ex]].
  
\bibitem{Alonso:2016oyd} 
  R.~Alonso, B.~Grinstein and J.~Martin Camalich,
  Phys.\ Rev.\ Lett.\  {\bf 118}, no. 8, 081802 (2017)
  [arXiv:1611.06676 [hep-ph]].

\bibitem{Akeroyd:2017mhr} 
  A.~G.~Akeroyd and C.~H.~Chen,
  Phys.\ Rev.\ D {\bf 96}, no. 7, 075011 (2017)
  [arXiv:1708.04072 [hep-ph]].
  
\bibitem{Sakaki:2013bfa} 
  Y.~Sakaki, M.~Tanaka, A.~Tayduganov and R.~Watanabe,
  Phys.\ Rev.\ D {\bf 88}, no. 9, 094012 (2013)
  [arXiv:1309.0301 [hep-ph]].

\bibitem{Alonso:2017ktd} 
  R.~Alonso, J.~Martin Camalich and S.~Westhoff,
  Phys.\ Rev.\ D {\bf 95}, no. 9, 093006 (2017)
  [arXiv:1702.02773 [hep-ph]].
  
\end{thebibliography}
\end{document}